\documentclass[sigconf]{acmart}
\usepackage{svg}
\usepackage{graphicx}

\usepackage{multirow}
\AtBeginDocument{%
  \providecommand\BibTeX{{%
    \normalfont B\kern-0.5em{\scshape i\kern-0.25em b}\kern-0.8em\TeX}}}

\copyrightyear{2025}
\acmYear{2025}
\setcopyright{acmlicensed}
\acmConference[IUI '25]{30th International Conference on Intelligent User Interfaces}{March 24--27, 2025}{Cagliari, Italy}
\acmBooktitle{30th International Conference on Intelligent User Interfaces (IUI '25), March 24--27, 2025, Cagliari, Italy}
\acmDOI{10.1145/3708359.3712098}
\acmISBN{979-8-4007-1306-4/25/03}

\begin{document}

\title{SkinGEN: an Explainable Dermatology Diagnosis-to-Generation Framework with Interactive Vision-Language Models}



\author{Bo Lin}
\authornote{Both authors contributed equally to this work.}
\affiliation{
  \institution{Zhejiang University}
  \institution{Binjiang Institute of Zhejiang University}
  \city{Hangzhou}
  \country{China}
}
\email{rainbowlin@zju.edu.cn}

\author{Yingjing Xu}
\authornotemark[1]
\affiliation{
  \institution{Zhejiang University}
  \city{Hangzhou}
  \country{China}
}
\email{poppyxu@zju.edu.cn}

\author{Xuanwen Bao}
\affiliation{
  \institution{The First Affiliated Hospital, Zhejiang University School of Medicine}
  \city{Hangzhou}
  \country{China}
}
\email{xuanwen.bao@zju.edu.cn}

\author{Zhou Zhao}
\affiliation{
  \institution{Zhejiang University}
  \institution{Shanghai Institute for Advanced Study of Zhejiang University}
  \city{Hangzhou}
  \country{China}
}
\email{zhaozhou@zju.edu.cn}


\author{Zhouyang Wang}
\authornotemark[2]
\affiliation{
  \institution{Hunan University}
  \city{Changsha}
  \country{China}
}
\email{wangzhouy@foxmail.com}



\author{Jianwei Yin}
\affiliation{
  \institution{Zhejiang University}
  \city{Hangzhou}
  \country{China}
}
\email{zjuyjw@cs.zju.edu.cn}

\begin{abstract}

With the continuous advancement of vision language models (VLMs) technology, remarkable research achievements have emerged in the dermatology field, the fourth most prevalent human disease category. However, despite these advancements, VLM still faces explainable problems to user in diagnosis due to the inherent complexity of dermatological conditions, existing tools offer relatively limited support for user comprehension. We propose SkinGEN, a diagnosis-to-generation framework that leverages the stable diffusion(SD) model to generate reference demonstrations from diagnosis results provided by VLM, thereby enhancing the visual explainability for users. Through extensive experiments with Low-Rank Adaptation (LoRA), we identify optimal strategies for skin condition image generation. We conduct a user study with 32 participants evaluating both the system performance and explainability. Results demonstrate that SkinGEN significantly improves users’ comprehension of VLM predictions and fosters increased trust in the diagnostic process. This work paves the way for more transparent and user-centric VLM applications in dermatology and beyond.

\end{abstract}



\maketitle
\begin{figure*}[ht]    
    \centering    
    \includegraphics[width=0.9\linewidth]{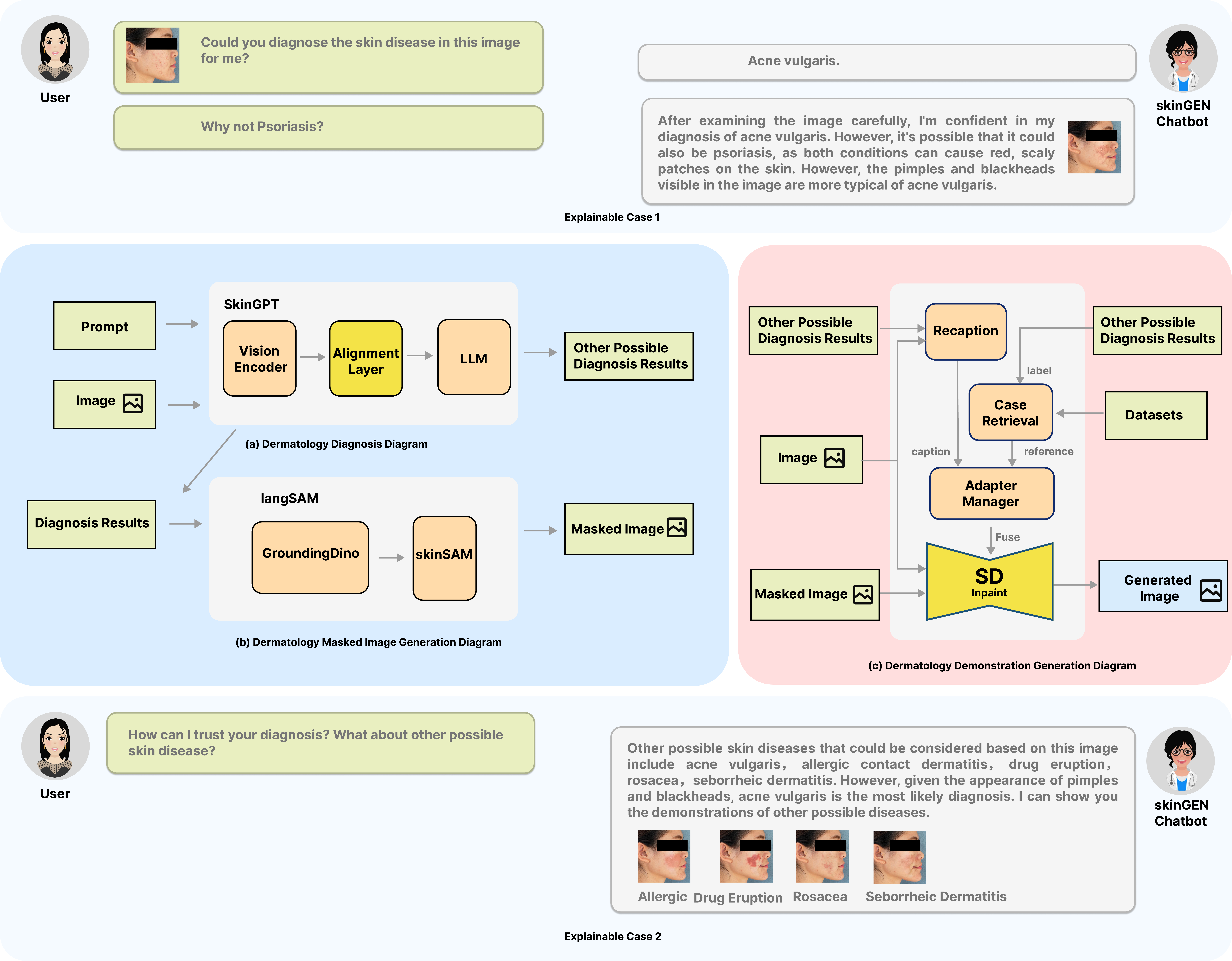}    
    \caption{SkinGEN Explainable Framework: (a) Dermatology Diagnosis Diagram: analyzes the user's image and provides a diagnosis along with potential alternatives.(b) Dermatology Masked Image Generation Diagram: generate a mask of the affected skin area (c) Dermatology Demonstration Generation Diagram: The Adapter Manager uses LoRA and/or Ip-adapter to generate visual examples of the diagnosed and possible conditions. \textbf{Case 1:} SkinGEN's diagnosis is questioned by the user. SkinGEN clarifies its reasoning and presents visual examples of similar conditions for comparison. \textbf{Case 2:} The user is unfamiliar with the diagnosis. SkinGEN provides visualizations of similar conditions to facilitate understanding and differentiation.} 
    \label{fig:solution}    
\end{figure*}

\section{Introduction}

In recent years, large language models (LLMs) \cite{zheng2023judging,touvron2023llama,wei2023larger,alpaca} and visual language models (VLMs) \cite{awadalla2023openflamingo,dai2023instructblip,li2023otter,liu2023visual,zhu2023minigpt,wang2023cogvlm} has witnessed remarkable and swift advancements. Several excellent VLMs such as miniGPT4 \cite{zhu2023minigpt}, openFlamingo \cite{awadalla2023openflamingo}, cogVLM \cite{wang2023cogvlm}, otter \cite{li2023otter} have shown extraordinary multi-model abilities of vision-language understanding and generation. In the healthcare field, VLMs have the potential to revolutionize the entire healthcare continuum by significantly optimizing and improving disease screening and diagnostic procedures, treatment planning, and post-treatment surveillance and care\cite{baldi2021deep}. These advancements offer a profound opportunity to transform healthcare practices, leading to improved patient outcomes and efficiency in healthcare delivery. Existing medical VLMs, such as MedViLL \cite{moon2022multi}, PubMedCLIP \cite{eslami2023pubmedclip}, LLaVa-Med \cite{li2024llava}, Med-Flamingo \cite{moor2023med}, XrayGPT \cite{thawkar2023xraygpt}, are tailored for visual question answering and report generation on extensive medical datasets. Adapting VLMs for medical visual question-answering is particularly noteworthy, empowering healthcare professionals to pose queries regarding medical images such as CT scans, MRIs, X-rays, and more \cite{hartsock2024vision}. Skin diseases are the fourth most common cause of all human diseases, affecting almost one-third of the world's population \cite{flohr2021putting}. SkinGPT-4 \cite{zhou2023skingpt}, an interactive dermatology diagnostic system trained on a vast repository of skin disease images, totaling 52,929 images encompassing both publicly available and proprietary sources, supplemented by clinical concepts and physicians' notes. Through fine-tuning, the model demonstrated comprehensive efficacy across the diagnostic process.
\newline
However, the reliability of these models has become a growing concern, particularly in real-world applications where users often find it difficult to trust the diagnostic results of large language models, as the visual foundation models operate as black boxes. Dermatological conditions may present similar symptoms in their early stages \cite{offidani2002general}, thus complicating user's cognition. The diagnosis of certain dermatological conditions is more challenging compared to other fields, owing to the potential similarity in symptoms and clinical manifestations despite potentially disparate etiologies, which poses significant challenges for patients lacking knowledge about dermatological knowledge. Solely focusing on a model’s predicted diagnosis lacks visual interpretation \cite{rao2023towards}. It is widely recognized that vision, as an intuitive and easily understandable mode of expression, plays a crucial role in enhancing user interpretability. This characteristic is particularly critical in the medical domain, where healthcare information often entails high levels of complexity and specialization, necessitating its communication to non-professionals in a manner that is both intuitive and comprehensible to ensure accurate understanding and effective utilization of the information. In this context, the emergence of image generation methods such as SD \cite{ramesh2022hierarchical} holds significant importance, offering new possibilities for privacy protection and the generation of illustrative explanations within medical scenarios. Prior work integrating SD \cite{akrout2023diffusion,dalmaz2022resvit,liang2022sketch,li2022neural} with skin-related research primarily focused on expanding datasets for dermatological conditions, without delving into the realm of user interaction. 
\newline
We introduce SkinGEN, an innovative diagnostic tool designed to enhance the interpretability of VLM through the utilization of the SD method. Users have the capability to upload images depicting dermatological conditions and pose corresponding medical inquiries. SkinGEN, in turn, provides tailored diagnoses or other medical suggestions. In instances where users harbor doubts or encounter confusion regarding the diagnosis, SkinGEN introduces a feature for demonstrating skin diseases, providing illustrative images depicting alternative dermatological diagnoses similar to the current diagnosis, thereby aiding users in distinguishing between them. Medical information can be transmitted and shared in a more secure and controlled manner through SD, while also enabling the creation of visually intuitive graphical representations. The SkinGEN framework comprises three diagrams: the dermatology diagnosis diagram, the dermatology masked image generation diagram, and the dermatology demonstration generation diagram. By investigating Low-Rank Adaptation (LoRA) within our framework, we established effective techniques for generating realistic and informative skin condition images, improving user comprehension and trust in VLM diagnoses. We recruited 32 participants for our user study. Through comparative experiments, we demonstrate that SkinGEN received positive recognition regarding perceived trust, ease of understanding, and cognitive effort, thereby validating its explainability. Furthermore, in the comprehensive evaluation of the system, participants also provided positive feedback, indicating that the system possessed attributes of being informative, useful, and easy to comprehend.
Our contributions are as follows:
\begin{enumerate}
\item  SkinGEN, innovatively uses both interactive VLMs and image generation to improve user understanding and trust. By visualizing the predicted skin condition and other possibilities, SkinGEN makes VLM diagnosis clearer and more reliable for users.
\item  Through extensive exploration of various training strategies and image synthesis methods, including fine-tuning SD with LoRA and incorporating both text and image prompts, we developed a highly effective solution for generating realistic and informative skin disease images.
\item User studies confirm that SkinGEN significantly improves user comprehension and trust in VLM diagnoses. This improvement is achieved by generating personalized visualizations of potential skin conditions directly from user-uploaded images, offering a clear and intuitive understanding of the diagnostic results while preserving user privacy.
\end{enumerate}

\vspace{-0.3cm}
\section{RELATED WORK}
\subsection{Image Generation}
Diffusion models, particularly Stable Diffusion(SD) \cite{rombach2022highresolution}, have revolutionized the field of image generation with their ability to synthesize high-quality images aligned with textual prompts. Trained on a massive dataset of images and text descriptions (LAION-5B) \cite{schuhmann2022laion5b}, SD leverages a latent diffusion process to progressively denoise an initial noise map into the desired image. This process can be further conditioned on various elements, including text prompts, class labels, or low-resolution images, enabling controlled and versatile image generation.
The desire for personalized image-generation experiences has driven the exploration of model customization techniques. Fine-tuning SD on domain-specific datasets with designated concept descriptors allows for tailoring the model to specific concepts or styles \cite{ruiz2023dreambooth, kumari2023multi}. This involves minimizing the original loss function of SD on the new data, enabling the model to learn and represent the unique features of the target concepts.
LoRA \cite{hu2021lora} has emerged as a powerful tool for enhancing the efficiency and effectiveness of model customization. By constraining the fine-tuning process to a low-rank subspace within the original parameter space of SD, LoRA significantly reduces the number of parameters that require updating while preserving the foundational knowledge of the pre-trained model. Building upon the success of text-to-image diffusion models like SD, research has explored efficient and controllable image generation methods, such as Ip-Adapter \cite{ye2023ip}, which leverages the power of both text and image prompts through a decoupled cross-attention mechanism.
\subsection{Medical VLM}
In recent years, significant advancements have been made in the field of LLMs  \cite{zheng2023judging,touvron2023llama,wei2023larger,alpaca}  and VLMs \cite{awadalla2023openflamingo,dai2023instructblip,li2023otter,liu2023visual,zhu2023minigpt,wang2023cogvlm}. MiniGPT4 \cite{zhu2023minigpt}, a generative visual-language model, trained through fine-tuning tasks on specialized datasets, leading to subsequent follow-up endeavors such as PatFig \cite{aubakirova2023patfig}, SkinGPT-4 \cite{zhou2023skingpt}, and ArtGPT-4 \cite{yuan2023artgpt}. These models are designed to address corresponding vision-language tasks across diverse domains. 
In the medical domain, vision language diagnostic models are regarded as an extremely promising direction for medical advancement, capable of addressing issues related to healthcare resource scarcity and the automation of intelligent diagnosis. Li etc proposed LLaVa-Med \cite{li2024llava}, an adaptation of the LLava \cite{liu2023visual}, specifically tailored for the medical domain through training on three standard biomedical visual question-answering datasets, which exhibits excellent multimodal conversational capability about a biomedical image. Micheal etc proposed Med-Flamingo \cite{moor2023med}, a multimodel few-shot learner that pre-trained on paired and interleaved medical image-text data from publications and textbooks based on OpenFlamingo-9B \cite{awadalla2023openflamingo}. These medical VLMs are fine-tuned on generative models using biomedical datasets, allowing the models to assimilate knowledge pertinent to the medical domain, thereby facilitating tasks such as medical diagnostics. The dataset utilized for these medical VLMs is mainly derived from in-vivo diagnostics, such as X-ray and CT scans. However, these medical models predominantly concentrate on the scenarios of in vivo diagnostics. In dermatology scenario, Zhou etc al. presented SkinGPT-4\cite{zhou2023skingpt}, which is the world's first interactive dermatology diagnostic system powered by an advanced visual language model MiniGPT-4. SkinGPT-4 was trained on an extensive collection of skin disease images(comprising 52,929 publicly available and proprietary images) along with clinical concepts and doctor's notes. The fine-tuned model shows a comprehensive performance in the diagnostic process, user comprehension enhancement, human-centered care, and healthcare equity promotion.
\subsection{Explainable AI}

To satisfy the stringent interpretability requirements of Explainable Artificial Intelligence(XAI), some prior works focus on data explanation so that humans can easily understand them\cite{wu2024usable,chen2021evaluating,welleck2022naturalprover,castillo2022chat}. For example, Chen et al. \cite{chen2021evaluating} provided explanatory comments to increase the readability and understandability of the generated code. Wekkeck et al. \cite{welleck2022naturalprover} proposed to explain math theorems by providing detailed derivations. However, these efforts aim to enhance user explainability in specific scenarios by leveraging the language generation capabilities of large language models. Visual information can enhance user interpretability, and SD possesses powerful visual generation capabilities. Therefore, utilizing the strong visual generation abilities of SD is an important approach to enhancing user interpretability.\\
Trust, alignment with clinical needs, and ethical deployment are critical components for successfully integrating these models into healthcare workflows \cite{hartsock2024vision}. The medical field prioritizes interpretability, while VLM's inherent lack of interpretability poses significant challenges to healthcare applications. Previous work in dermatoscopic synthetic data generation through SD to mitigate challenges associated with limited labeled datasets, thereby facilitating more effective model training \cite{akrout2023diffusion,dalmaz2022resvit,liang2022sketch,li2022neural}. Akrout et al. \cite{akrout2023diffusion} proposed text to image synthesis method for generating high-quality synthetic images of macroscopic skin diseases. Farooq et al. \cite{farooq2024derm} proposed Derm-T2IM that uses natural language text prompts as input and produces high-quality malignant and benign lesion imaging data as the output.

\section{Dermatology Demonstration Generation Method}
\subsection{Dataset}
\begin{figure}[!h]
    \centering
    \includegraphics[width=0.8\columnwidth]{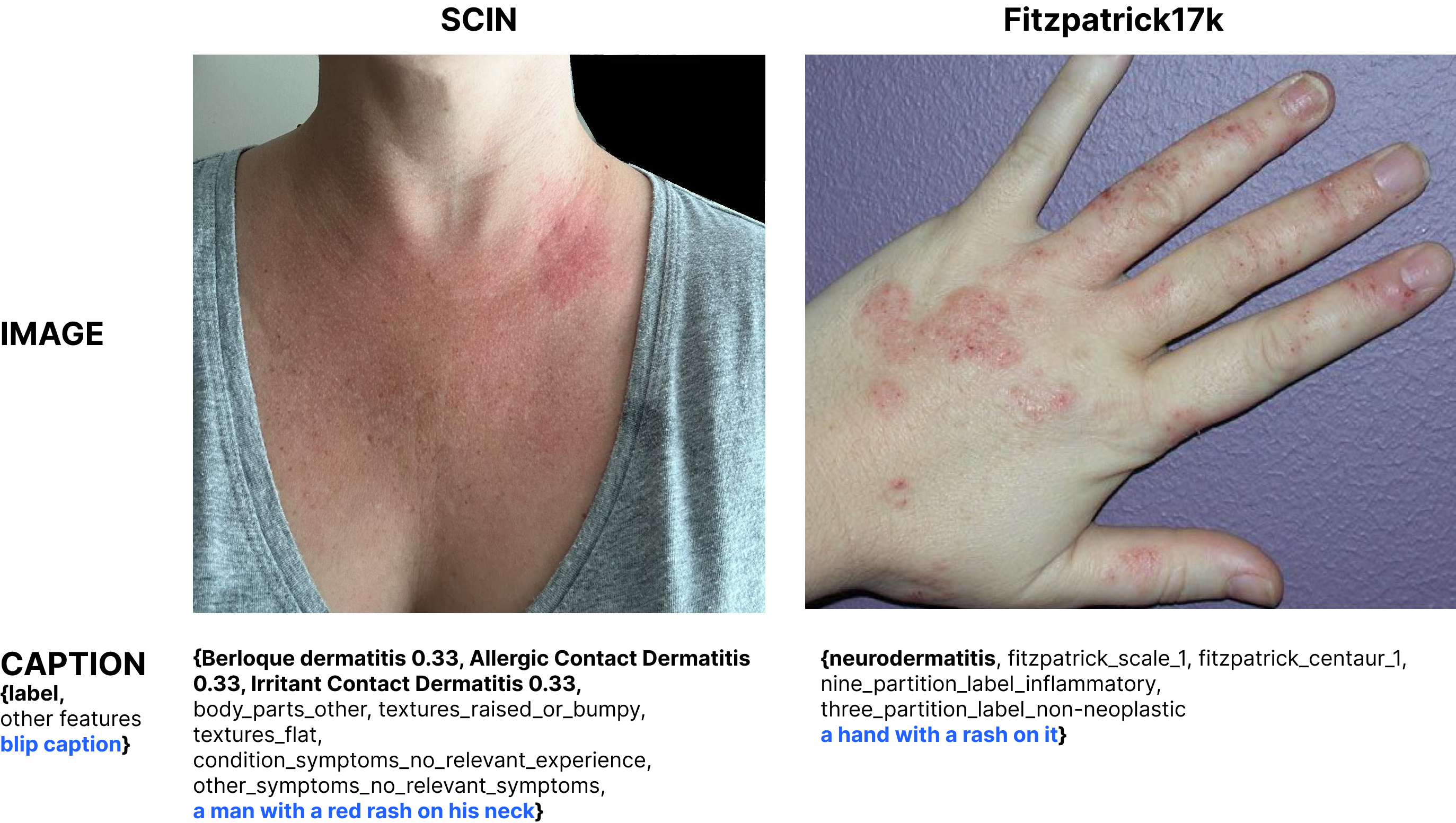}
    \caption{Example of an image-caption pair from the training data, where the caption is augmented with a detailed description generated by Blip2 \cite{li2023blip}.}
    \label{fig:data_example}
    \vspace{-0.3cm} 
\end{figure}

This study utilizes two relevant datasets for training the skin disease generation model: Fitzpatrick17k \cite{groh2021evaluating} and the recently released SCIN Dataset \cite{ward2024crowdsourcing}. Fitzpatrick17k comprises 16,577 clinical images encompassing 114 skin conditions annotated by dermatologists. It also includes discrete labels such as the Fitzpatrick scale, which describes various aspects of skin disease conditions. In contrast, the SCIN Dataset is crowdsourced, collected from 5,000 volunteers, and contains over 10,000 images. SCIN employs a weighted condition labeling system, where a case may be associated with up to 3 skin diseases (conditions).
Both datasets exhibit an imbalanced distribution of labels. Fitzpatrick17k's most frequent condition is psoriasis' with 653 cases, while the least frequent ispilomatricoma' with only 52 cases. SCIN demonstrates a similar imbalance. To address this and prepare the data for model training, we sampled three subsets of varying scales: 5-shot, 30-shot, and all. The 5-shot and 30-shot subsets contain random samples across all labels, while the "all" dataset encompasses the entirety of the data.
For the training data, we extracted primary skin condition labels and associated features from image-caption pairs, as illustrated in Figure \ref{fig:data_example}. We then employed the Blip2 model \cite{li2023blip} to augment the captions with detailed descriptions while also retaining an original dataset with only the extracted labels as captions for comparison purposes.

As Table \ref{table:LoRA_date} illustrates, this data preprocessing resulted in 12 subsets (3 scales x 2 datasets x 2 captioning methods), allowing us to investigate the following: (1) \textbf{Labeling Method:} We compare the effectiveness of single-label annotations (Fitzpatrick17k) versus multi-weight labels (SCIN). (2) \textbf{Scaling Effect:} We analyze the performance across 5-shot, 30-shot, and all datasets to understand the impact of the data scale. (3) \textbf{Blip Captioning:} We assess whether the addition of detailed descriptions through Blip2 improves model performance.

\subsection{Model Training}
\textbf{Adapter-based Fine-tuning.} To facilitate skin disease image generation conditioned on specific skin conditions, we employed two adapter-based fine-tuning techniques: LoRA \cite{hu2021lora} and Image Prompt Adapters (Ip-adapter) \cite{ye2023ip}. Both methods introduce additional trainable layers that integrate seamlessly into the U-Net \cite{ronneberger2015u} and Cross-Attention \cite{vaswani2017attention} layers of the SD v1.5 \cite{ramesh2022hierarchical} image generation pipeline. 

\textbf{LoRA Configuration.}  To investigate the impact of training data volume on model performance, we trained LoRA models using three datasets with varying numbers of samples, as detailed in Table~\ref{table:LoRA_date}. We adjusted the LoRA dimension (dim) according to the dataset size while maintaining consistent hyperparameters across all experiments. These hyperparameters included 20 epochs, a batch size of 2, the AdamW optimizer with 8-bit precision \cite{loshchilov2017decoupled}, a learning rate of $1e^{-4}$, a text encoder learning rate of $5e^{-5}$, mixed precision (FP16), and dataset image resolution resized to $512 \times 512$ pixels. 

\begin{table}[h]
 \vspace{-0.3cm} 
\caption{Table of dataset and LoRA parameters}
 \vspace{-0.4cm} 
\label{table:LoRA_date}
\begin{tabular}{r c c}  
\toprule
\textbf{Dataset} & \textbf{Image counts} & \textbf{LoRA dim} \\
\midrule
f17k-5-shot & \multirow{2}{*}{570} & \multirow{2}{*}{32} \\ 
f17k-5-shot-blip & & \\ 
f17k-30-shot & \multirow{2}{*}{3470} & \multirow{2}{*}{64} \\
f17k-30-shot-blip & & \\
f17k-All & \multirow{2}{*}{16576} & \multirow{2}{*}{128} \\
f17k-All-blip & & \\
\midrule 
SCIN-5-shot & \multirow{2}{*}{1547} & \multirow{2}{*}{32} \\ 
SCIN-5-shot-blip & & \\ 
SCIN-30-shot & \multirow{2}{*}{3341} & \multirow{2}{*}{64} \\
SCIN-30-shot-blip & & \\
SCIN-All & \multirow{2}{*}{7798} & \multirow{2}{*}{128} \\
SCIN-All-blip & & \\
\bottomrule
\end{tabular}
 \vspace{-0.3cm} 
\end{table}
\textbf{Ip-adapter.} Due to limited computational resources, we initially trained LoRA models using all 12 datasets. We conducted experiments to identify the top-performing LoRA model and its corresponding dataset. Subsequently, we trained an Ip-adapter model using the dataset selected from the LoRA training phase. This allowed us to evaluate the skin disease image generation performance of the Ip-adapter model in comparison to the LoRA approach. 


\subsection{LoRA Evaluation}
\textbf{Quantitative Analysis.}  
The test dataset consists of 100 image-caption pairs sampled from the original training dataset. To evaluate the quality and effectiveness of generated skin disease images, we employed three key metrics to measure the similarity between the original and generated image (using only the caption paired with the original image): (1) \textbf{CLIP score} \cite{radford2021learning} for semantic similarity, (2) \textbf{DINOv2 score} \cite{oquab2023dinov2} for structural and perceptual quality, and (3) \textbf{Mean Squared Error (MSE)} for pixel-level fidelity. We calculated the \textbf{BLIP Gain} as the improvement from using BLIP-generated captions (averaged across 5-shot, 30-shot, and all-shot models) and the \textbf{Scaling Effect} as the difference between the full dataset (\texttt{all}) and 30-shot models.  

As shown in Table~\ref{table:evaluation_metrics}, the incorporation of BLIP captions yields mixed improvements. For the Fitzpatrick17k dataset, BLIP captions marginally improve CLIP scores (+0.02) and MSE (+0.01), while DINOv2 scores remain unchanged. In contrast, the SCIN dataset demonstrates stronger gains, with CLIP scores improving by +0.08, MSE by +0.05, and DINOv2 scores showing a minor increase (+0.03). Analysis of scaling effects reveals divergent trends between datasets. Scaling to the full Fitzpatrick17k dataset degrades CLIP ($-0.05$) and DINOv2 ($-0.05$) scores compared to 30-shot models, despite improving MSE (+0.08), suggesting potential overfitting or data imbalance in larger training sets. For the SCIN dataset, scaling slightly reduces CLIP ($-0.02$) and DINOv2 ($-0.01$) scores, while MSE worsens ($-0.04$), indicating challenges in maintaining fidelity with increased data.  

These results highlight inconsistencies in the benefits of BLIP captions and dataset scaling. While BLIP marginally enhances semantic alignment (CLIP) and pixel fidelity (MSE), its impact on structural quality (DINOv2) is negligible or dataset-dependent. The degradation of CLIP/DINOv2 scores with larger datasets underscores the need for refined training strategies, such as balancing class distributions or regularization, to fully leverage the potential of large-scale data in medical image generation.  


\begin{table}[h]
\caption{Evaluation Metrics for Trained LoRA Models}
\vspace{-0.4cm}
\label{table:evaluation_metrics}
\begin{tabular}{r l l l}
\toprule
\textbf{Model} & \textbf{CLIP} $\uparrow$ & \textbf{DINOv2} $\uparrow$ & \textbf{MSE} $\downarrow$ \\
\midrule
0-shot & 0.61 & 0.69 & 2.09 \\
f17k\_5shot & 0.76 & 0.81 & 1.31 \\
f17k\_5shot\_blip & 0.77 & 0.82 & 1.32 \\
f17k\_30shot & 0.76 & 0.82 & 1.33 \\
f17k\_30shot\_blip & 0.76 & 0.82 & 1.31 \\
f17k\_all & 0.71 & 0.77 & 1.25 \\
f17k\_all\_blip & 0.72 & 0.76 & 1.25 \\
\hline
\textbf{BLIP Gain} & +0.02 & +0.00 & +0.01 \\
\textbf{Scaling Effect} & -0.05 & -0.05 & +0.08 \\
\hline
0-shot & 0.63 & 0.71 & 2.02 \\
SCIN\_5shot & 0.71 & 0.76 & 1.64 \\
SCIN\_5shot\_blip & 0.75 & 0.77 & 1.61 \\
SCIN\_30shot & 0.74 & 0.76 & 1.58 \\
SCIN\_30shot\_blip & 0.76 & 0.77 & 1.53 \\
SCIN\_all & 0.72 & 0.75 & 1.62 \\
SCIN\_all\_blip & 0.74 & 0.76 & 1.54 \\
\hline
\textbf{BLIP Gain} & +0.08 & +0.03 & +0.05 \\
\textbf{Scaling Effect} & -0.02 & -0.01 & -0.04 \\
\bottomrule
\end{tabular}
\end{table}

\begin{figure}[!h]
    \centering
    \includegraphics[width=0.9\columnwidth]{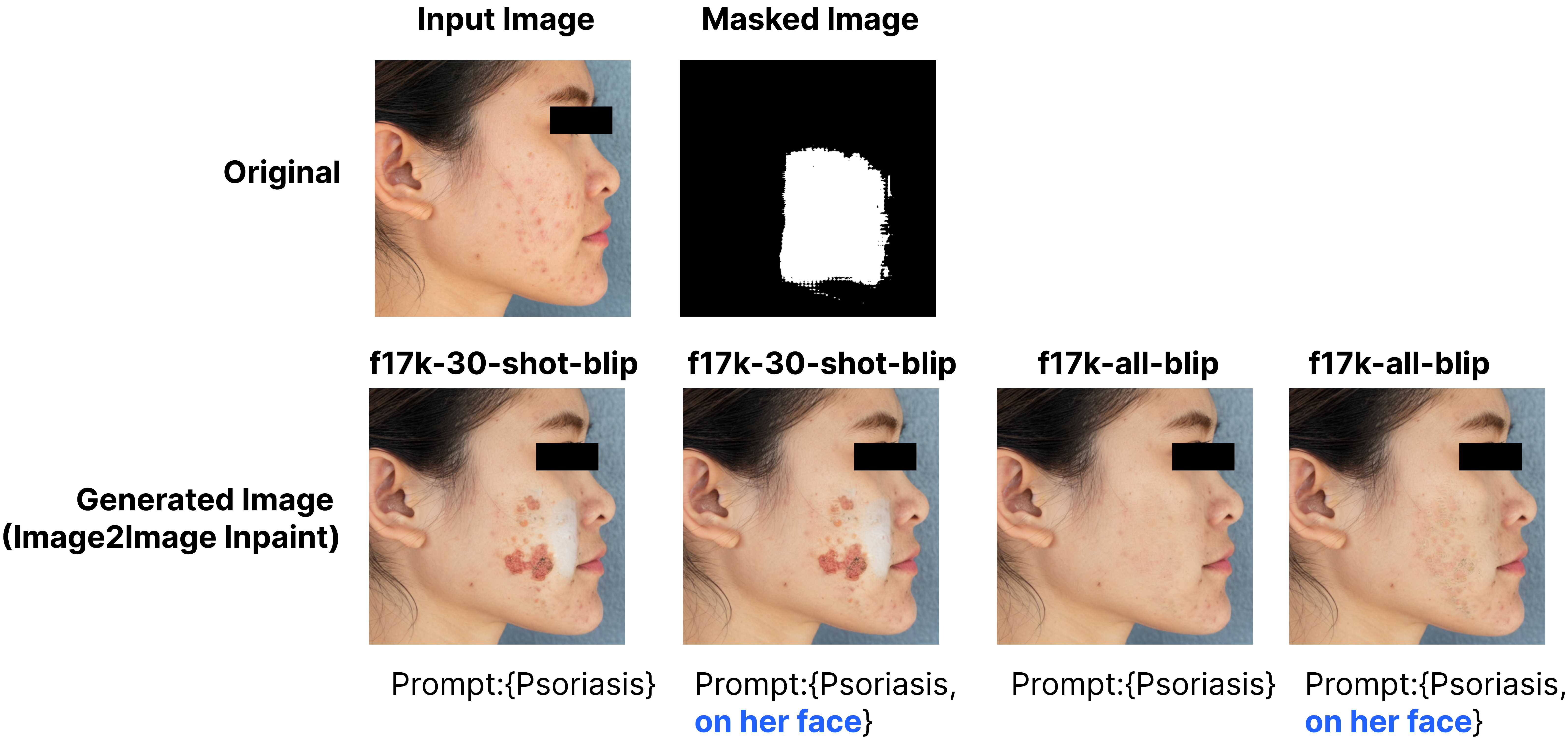}
\caption{Comparison of skin disease image generation using LoRA models with and without blip captions, showcasing the influence of textual descriptions and training data size on the models' semantic understanding of disease presentation.}
\label{fig:semantic_understanding}
\end{figure}

\textbf{Qualitative Analysis of Semantic Understanding.} To further explore the influence of blip captions on the semantic understanding of our trained models, we conducted a qualitative analysis of generated skin disease images. Figure~\ref{fig:semantic_understanding} showcases the results of this analysis. We observed that LoRA models trained with longer training steps (i.e., using larger datasets) exhibited evidence of developing a more nuanced understanding of skin conditions and their presentation.  For instance, when generating images using only the label "Psoriasis" as the caption, both the "f17k-30shot-blip" and "fk17k-all-blip" LoRA models produced similar outcomes, depicting psoriasis symptoms on the body. However, when the caption was augmented with the additional description "on her face," the models' behavior diverged.  The "f17k-all-blip" LoRA model, trained on the full Fitzpatrick17k dataset, seemingly learned from the data distribution that psoriasis is less likely to occur on the face. Consequently, it generated an image with clear skin on the face, aligning with the typical presentation of the condition. In contrast, the "f17k-30shot-blip" model, trained on a smaller subset of the data, adhered more directly to the provided caption and generated psoriasis symptoms on the girl's face, albeit with a milder appearance.  These observations suggest that larger models, exposed to a broader range of examples, develop a more comprehensive understanding of disease characteristics and potential variations, enabling them to generate images that are both realistic and consistent with the provided captions.

\begin{figure}[!h]
    \centering
    \includegraphics[width=0.9\columnwidth]{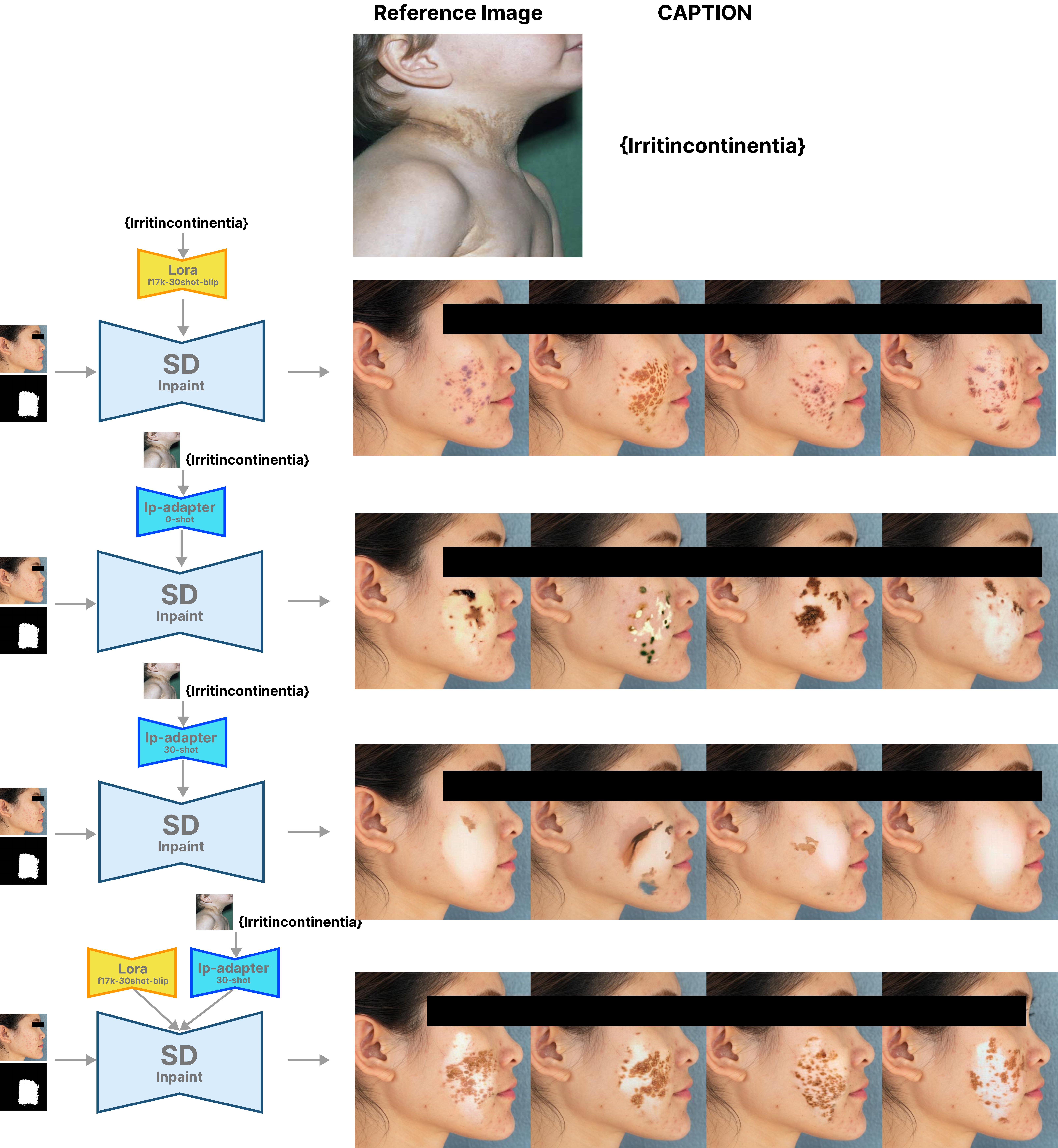}
    \caption{Comparison of Skin Disease Image Generation using Different Adapter Configurations. The figure showcases example outputs from four experimental conditions: (1) LoRA model, (2) 0-shot Ip-adapter, (3) fine-tuned Ip-adapter, and (4) fused Ip-adapter and LoRA model. The results highlight the advantages of adapter fusion in generating disease patterns that closely resemble the reference images.}
    \label{fig:ip_adapter}
\end{figure}

\textbf{Evaluation of Ip-adapter and Adapter Fusion.} To assess the efficacy of Ip-adapter and the potential benefits of adapter fusion, we designed a comparative experiment with four conditions, as visualized in Figure~\ref{fig:ip_adapter}: (1) LoRA (f17k\_30shot\_blip), (2) Ip-adapter (0-shot), (3) Ip-adapter (30-shot) fine-tuned on the f17k\_30shot\_blip dataset for 20,000 steps, and (4) Ip-adapter (30-shot) + LoRA (30-shot) representing adapter fusion. Our findings indicate that the fused adapter configuration (condition 4) yielded the most favorable outcomes, generating disease patterns that closely resembled the reference images. Notably, the 0-shot Ip-adapter demonstrated an inherent ability to grasp the task and synthesize skin disease patterns, albeit with less accuracy compared to the fine-tuned and fused models.  Surprisingly, the fine-tuned Ip-adapter (30-shot) produced the least desirable results, suggesting potential challenges related to overfitting or optimization during the fine-tuning process. 

\section{Implementation}
\subsection{Explainable Framework}
SkinGEN is a diagnosis assistant in the field of dermatology, which provides a dermatology diagnosis-to-generation solution to increase VLM's visual explainability to users. For skin diseases that are difficult to distinguish, it utilizes SD to provide visual demonstrations for similar skin diseases to help patients better distinguish between them. As illustrated in Fig. \ref{fig:solution}, the solution consists of three diagrams: dermatology diagnosis diagram, dermatology masked image generation diagram, and dermatology demonstration generation diagram. In the dermatology diagnosis diagram, we employ SkinGPT-4 \cite{zhou2023skingpt} to diagnose the uploaded image, obtaining diagnostic outcomes and other potential skin disease results. In the dermatology masked image generation diagram, we first utilize the GroundingDINO \cite{liu2023grounding}  model to identify the location of skin disease within the image based on the prompt of skin conditions. Subsequently, the skinSAM model \cite{hu2023skinsam} is employed to segment the regions affected by skin diseases, generating a masked image. In the dermatology demonstration generation diagram, Our exploration of adapter methods led to the development of a robust image generation approach. In the subsequent sections, we will delve into the technical details of implementing these three diagrams.
\subsection{Dermatology Diagnosis}

SkinGPT-4 is an interactive system in the field of dermatology designed to provide a natural language-based diagnosis of skin disease images \cite{zhou2023skingpt}. It was trained on an extensive collection of skin disease images(comprising 52,929 publicly available and proprietary images) along with clinical concepts and doctors' notes. The architecture of SkinGPT-4 comprises three modules: the visual encoder, the projection layer, and an advanced large language decoder. The visual encoder includes the Vision Transformer(VIT)\cite{dosovitskiy2020image} and Q-Transformer\cite{chebotar2023q}, which can encode the input image into image embedding account for the image's context. The function of the alignment layer is to align the semantics of the text space with the semantics of the image space. The SkinGPT-4 utilizes the Vicuna\cite{wei2023larger} and Llama\cite{touvron2023llama} as the language decoder, which can perform a wide range of complex linguistic tasks.

\begin{figure}[!h]    
    \centering    
    \includegraphics[width=0.65\columnwidth]{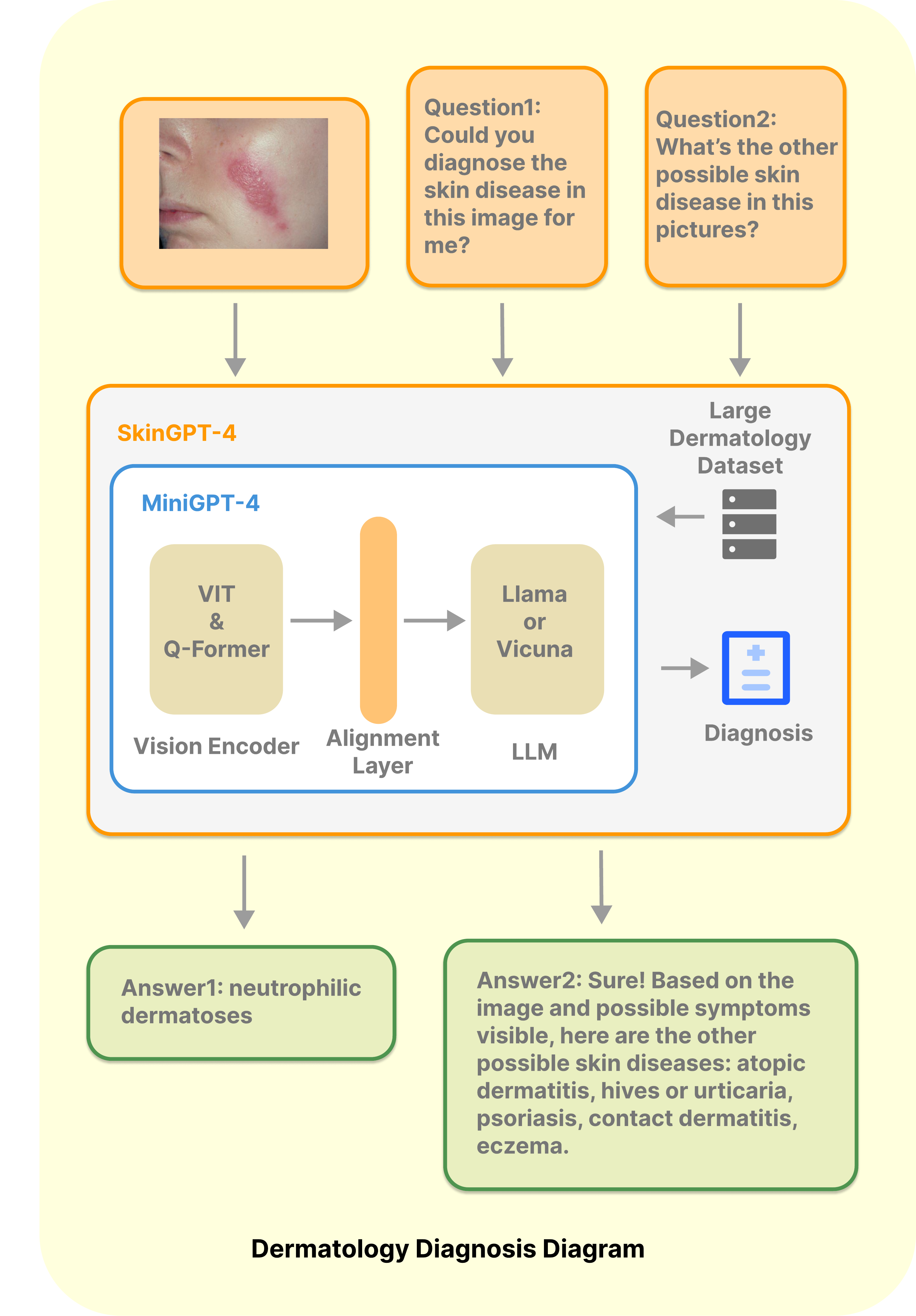}    
    \caption{Dermatology Diagnosis with SkinGPT-4 \cite{zhou2023skingpt}.}    
    \label{fig:diagnose_pipeline}    
     \vspace{-0.3cm} 
\end{figure}

We utilize SkinGPT-4 to accept dermatology images and questions from users. Upon receiving a query from the user, SkinGPT-4 provides corresponding answers. Fig. \ref{fig:solution} shows two examples of SkinGPT-4, it can be observed that SkinGPT-4 possesses knowledge and understanding of dermatological conditions, enabling it to assist patients in making preliminary diagnoses and providing relevant suggestions. Due to the inherent complexity in diagnosing skin diseases, SkinGPT-4 may not always provide entirely accurate diagnostic outcomes; however, it is also capable of generating other possible diagnoses for the given image. The pipeline of skin disease is in Fig. \ref{fig:diagnose_pipeline}. The user uploads an image of skin disease and poses questions, the process by which SkinGPT-4 handles user prompts can be divided into two steps: encoding and decoding. In the process of encoding, the visual encoder extracts vital features and generates an embedding of the image based on the features. The alignment layer synchronizes the visual information and natural language, thus the visual embedding is transformed into an embedding with textual semantics. The input prompt is tokenized by the tokenizer and then it concatenates with the transformed visual embedding. In the processing of decoding, we input the concatenated embedding into Vicuna, which generates the text-based diagnosis. To obtain both the diagnostic result and other possible skin results for the current dermatological image, we designed two tasks as shown in Fig. \ref{fig:diagnose_pipeline}. The first task aims to obtain the diagnostic result of the skin disease in the input image, the prompt designed for the first task is "Could you diagnose the skin disease in this image for me?". We aim to obtain the diagnostic result of a skin disease, such as "acne". The second task aims to identify other possible skin diseases, the prompt designed for the second task is "What's the other possible skin disease in this picture?". We expect to receive a list containing possible skin disease outcomes, for instance, ["Atopic dermatitis", "Hives or urticaria", "Psoriasis", "Contact dermatitis", "Eczema"].
\subsection{Dermatology Masked Image Generation}
\begin{figure}[!h]    
    \centering    
    \includegraphics[width=0.65\columnwidth]{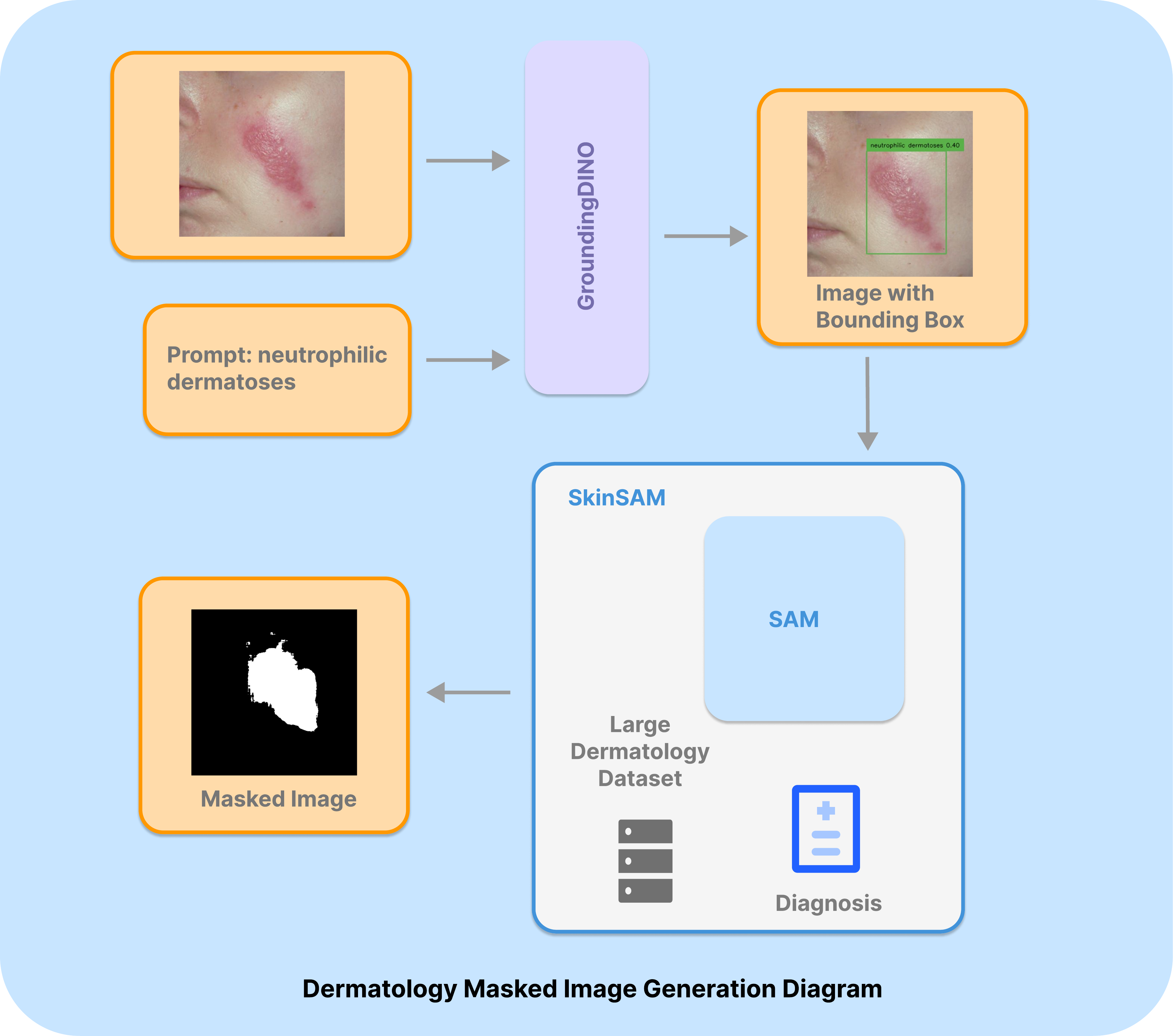}    
    \caption{Dermatology masked image generation diagram with object identification and mask generation.}    
    \label{fig:mask_pipeline}    
     \vspace{-0.3cm} 
\end{figure}

To generate masked images of skin disease, we apply the \href{https://github.com/luca-medeiros/lang-segment-anything}{lang-segment-anything(lang-SAM)} pipeline. Different from the global segmentation tasks the segment anything model(SAM) \cite{kirillov2023segany} provided, the lang-SAM can identify and segment the dermatological area by prompt in the skin image. Fig. \ref{fig:mask_pipeline} illustrates the process of dermatology masked image generation. First, we use the GroudingDino model to identify the skin disease region in the uploaded image. GroundingDINO \cite{liu2023grounding} is an object identification model that requires a prompt input, when the prompt is set to a specific skin disease, the GroundingDINO is capable of returning the bounding box indicating the location of the specific skin disease within the current image. After obtaining the bounding box position of the skin disease, we proceed to generate the mask of this image. We input the current skin disease image along with the bounding box information into the skinSAM, it subsequently returns a masked image of the skin disease.

It is noteworthy that the SAM is trained on the SA-1B dataset of 11 million images and 1.1 billion masks, which can understand the visual concepts deeply. To adapt the SAM for skin cancer segmentation tasks better, we employed the skinSAM \cite{hu2023skinsam}, which is a fine-tuned version of the SAM validated on HAM10000 dataset \cite{tschandl2018ham10000} specifically designed for skin diseases which includes 10015 dermatoscopic images. We downloaded the weights of the \href{https://huggingface.co/ahishamm/skinsam}{skinSAM} from Huggingface for our use. Figure \ref{fig:mask_pipeline} provides masked image examples generated by lang-SAM.
\begin{figure}[!h]  
    \centering  
    \includegraphics[width=0.5\columnwidth]{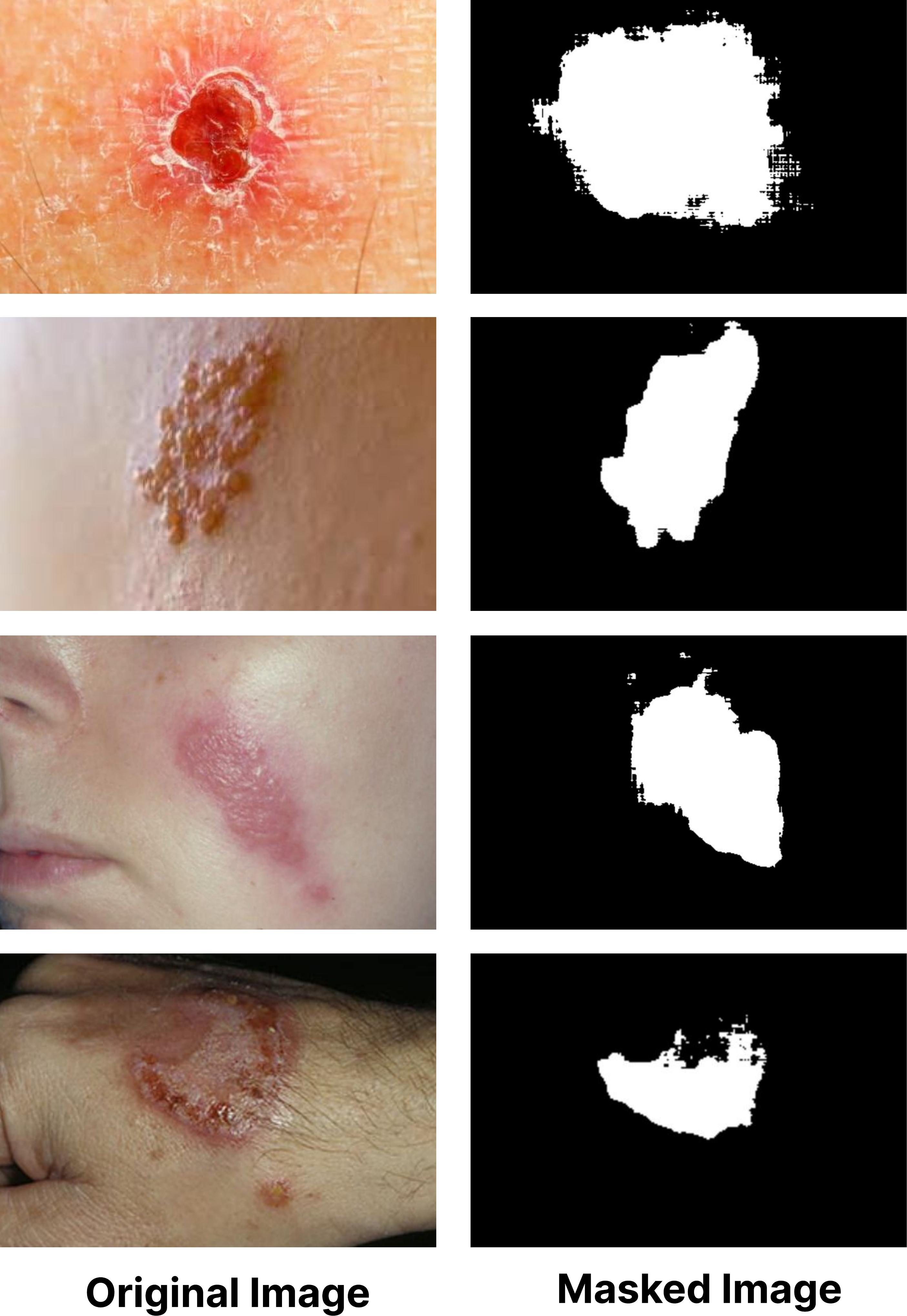}  
    \caption{Example of masked images generated by lang-sam from dermatological images.}  
    \label{fig:mylabel}  
    \vspace{-0.3cm} 
\end{figure}


\subsection{Dermatology Image Generation}

\begin{figure}[!h]
    \centering
    \includegraphics[width=0.65\columnwidth]{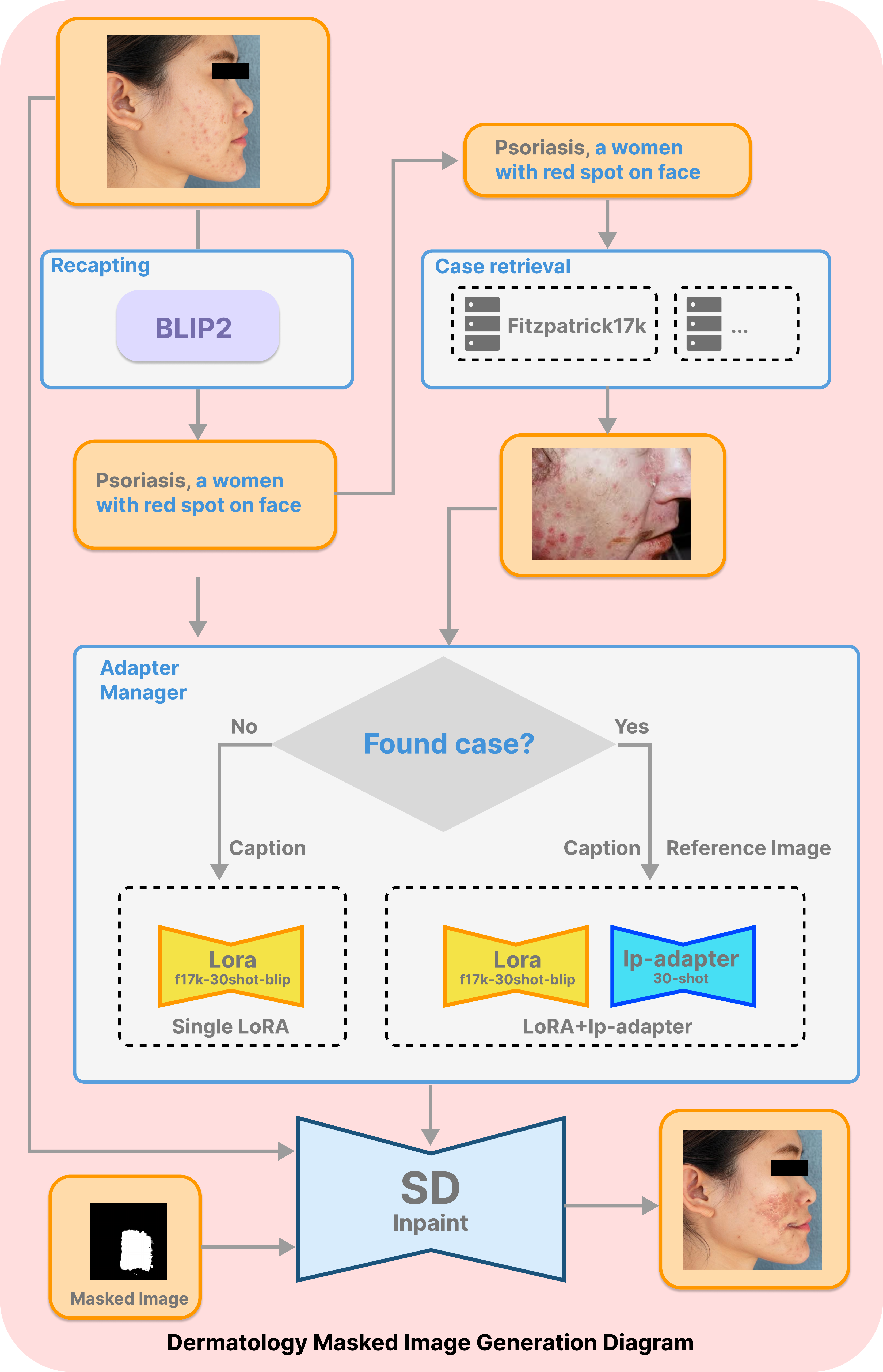}
    \caption{Dermatology image generation diagram with recaptioning, case retrieval, and adapter management modules.}
    \label{fig:image_gen_diagram}
\end{figure}

Building upon the generative model experiments in Section 3, we propose a tailored image generation diagram for dermatology applications, addressing the challenge of generating skin disease images based on user-provided input. The pipeline comprises three key components: (1) \textbf{Recaptioning:} We employ BLIP2 \cite{li2023blip} to automatically generate descriptive captions for user-provided images. This step enhances the input information and provides a more informative context for subsequent stages. (2) \textbf{Case Retrieval:} A case retrieval module searches for relevant skin disease cases within a curated database based on the input image label and BLIP2-generated caption. This module aims to identify existing cases with similar characteristics to guide the image generation process. (3) \textbf{Adapter Manager:} This module dynamically selects the appropriate image generation strategy based on the case retrieval results. If relevant cases are found, we utilize a combination of LoRA and IP-Adapter to leverage both textual and visual information from the retrieved cases. In the absence of similar cases, we employ LoRA with a standard text-to-image generation approach using the BLIP2 caption as input. Figure \ref{fig:image_gen_diagram} illustrates the overall architecture of the proposed dermatology image generation pipeline.

\section{User Study}

\subsection{Experiment Design}

To evaluate the effectiveness of different skin disease explanation systems, we conducted an online user study with 32 participants recruited through social media. Regarding participant backgrounds, the majority are university students; Table \ref{tab:background} presents the statistics. All participants had agreed to collect data and record the experiment process. The study employed a within-subjects design where each participant interacted with three distinct systems: (1) \textbf{System 1 (SkinGPT): } This system provided plain text explanations of skin conditions generated by the SkinGPT. (2) \textbf{System 2 (Reference Case Retrieval): } This system retrieved and presented visually similar cases from a database of skin disease images along with their corresponding diagnoses. (3) \textbf{System 3 (SkinGEN): } Our proposed SkinGEN system, diagnosed skin conditions based on user-uploaded images and generated personalized visualizations of the identified diseases.

The experiment was implemented using a multi-modal chatbot application built with Gradio \cite{abid2019gradio}. Participants engaged in conversations with the chatbot, providing information about their skin concerns and receiving explanations from each system in a randomized order. 

\begin{table}[H]
\centering
\vspace{-0.4cm}
\caption{Gender and Medical Background Distribution}
\vspace{-0.4cm}
\begin{tabular}{llcc}
\toprule
 & \textbf{Option} & \textbf{Count} & \textbf{Percentage} \\
\midrule
\multirow{3}{*}{\textbf{Gender Distribution}} & Male & 22 & 68.75\% \\
 & Female & 10 & 31.25\% \\
\midrule
\multirow{3}{*}{\textbf{Medical Background}} & Yes & 12 & 37.5\% \\
 & No & 20 & 62.5\% \\
\bottomrule
\end{tabular}
\vspace{-0.4cm}
\label{tab:background}
\end{table}

\subsection{Evaluation of Chatbot Explainability}
To assess user perceptions of the chatbot system's explainability, we employed established trust-related metrics inspired by prior work \cite{pu2006trust,2018Explaining}. These metrics focus on evaluating both system and user interface explainability through the lens of user trust. Following interactions with each system, participants completed a questionnaire designed to measure three key constructs: (2) \textbf{Perceived Trust:} The extent to which users felt they could rely on the system's information and recommendations. (2) \textbf{Ease of Understanding:} The clarity and comprehensibility of the system's explanations and conversational flow. (3) \textbf{Cognitive Effort:} The mental effort required by users to understand and process the information provided by the system.

All participants engaged in interactions with each of the three chatbot systems in a randomized order. During each system test, participants selected a skin disease image online (restricted to 114 labels within the Fitzpatrick17k dataset) and input it into the SkinGEN user interface to initiate a conversation with the chatbot. After each interaction, users rated their experience based on the questions presented in Table \ref{table:questionnaire_repeated}.
Participants rated their experiences using a 5-point Likert scale, where 1 indicated 'Strongly Disagree' and 5 indicated 'Strongly Agree,' to assess various aspects of chatbot explainability, including perceived trust, ease of understanding, and cognitive effort.
\vspace{-0.3cm} 
\begin{table}[h]
  \caption{System Explainability Evaluation}
  \vspace{-0.4cm} 
  \label{table:questionnaire_repeated}
  \centering
  \begin{tabular}{p{0.9\columnwidth}}
    \toprule
    \textbf{Question} \\
    \midrule
    Repeated measure (3 conditions)\\
    \midrule
    Perceived Trust: I can trust the system. \\
    Ease of Understanding: The conversation with the system is easy to understand. \\
    Cognitive Effort: I easily found the information I was asking for. \\
    \midrule
    Rated once\\
    \midrule
    SkinGEN’s diagnosis is correct or relevant. \\
    The description provided by SkinGEN is informative.\\
    The suggestions offered by SkinGEN are useful.  \\
    I would be willing to use SkinGEN in the future.  \\
    The generated skin disease image looks realistic. \\
    I find SkinGEN to be a useful system. \\
    \bottomrule
  \end{tabular}
  \vspace{-0.3cm} 
\end{table}
\subsection{Result}
\begin{figure}
    \centering
    \includegraphics[width=0.45\textwidth]{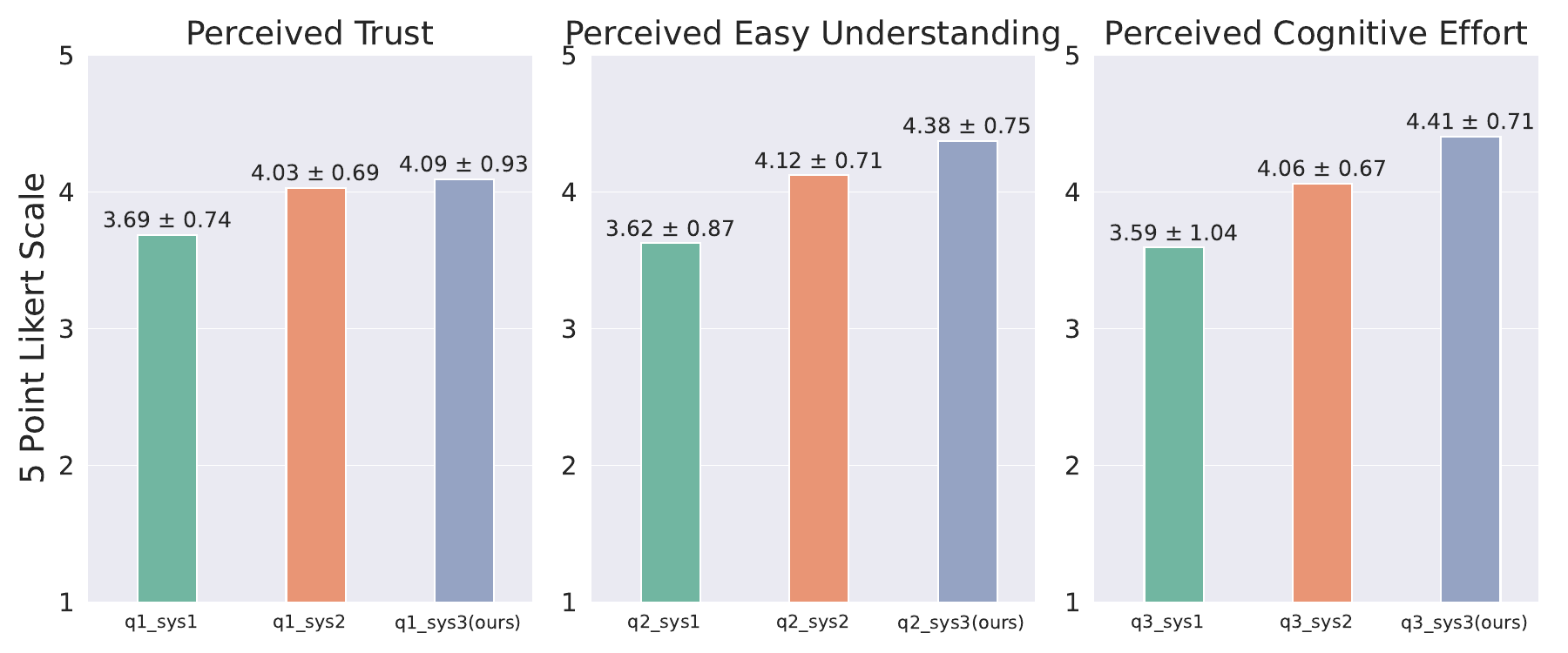}
    \caption{Perceived explainability ratings (Mean ± SD) for the three explanation systems: SkinGPT (sys1), SkinGPT + reference case retrieval (sys2), and SkinGEN (sys3, ours).}
    \label{fig:trust_score}
    \vspace{-0.4cm} 
\end{figure}

\textbf{Explainability} The user study results, as visualized in Figure~\ref{fig:trust_score}, provide insights into user perceptions of trust, ease of understanding, and cognitive effort associated with different skin disease explanation systems. System 1, which represents the plain SkinGPT explanation, achieved a moderate level of perceived trust and ease of understanding. System 2, based on finding reference cases in a database, showed slightly higher scores in both categories. Notably, System 3 (SkinGEN), our proposed method for diagnosis and skin disease image generation based on user-uploaded images, outperformed both baseline systems across all three metrics.  Users perceived SkinGEN explanations as significantly more trustworthy, easier to understand, and requiring less cognitive effort to interpret. This suggests that the ability to generate personalized visualizations of skin conditions based on individual cases resonates with users and enhances their comprehension of the provided information. 


\textbf{System Performance.} The results of the user questionnaire, summarized in Table~\ref{table:questionnaire_results}, demonstrate positive user perceptions of SkinGEN's performance. Participants generally agreed that SkinGEN's diagnoses were accurate or relevant (mean = 4.16), the provided descriptions were informative (mean = 4.41), and the suggestions offered were useful (mean = 4.31). Furthermore, users expressed a willingness to utilize SkinGEN in the future (mean = 4.31) and perceived the system as useful overall (mean = 4.38). The visual fidelity of the generated skin disease images also received positive feedback (mean = 4.16), indicating that users found the images to be realistic. These findings suggest that SkinGEN effectively addresses user needs in understanding and visualizing skin conditions, fostering trust and confidence in the system's capabilities. 
\vspace{-0.3cm} 
\begin{table}[h]
\caption{Questionnaire Results: Mean and Standard Deviation}
\vspace{-0.4cm} 
\label{table:questionnaire_results}
\begin{tabular}{p{6cm} p{1.5cm}}
\toprule
\textbf{Question} & \textbf{Mean ± SD} \\
\midrule
SkinGEN’s diagnosis is correct or relevant. & 4.16 ± 0.63 \\
The description provided by SkinGEN is informative. & 4.41 ± 0.61 \\
The suggestions offered by SkinGEN are useful. & 4.31 ± 0.78 \\
I would be willing to use SkinGEN in the future. & 4.31 ± 0.78 \\
The generated skin disease image looks realistic. & 4.16 ± 0.77 \\
I find SkinGEN to be a useful system. & 4.38 ± 0.75 \\
\bottomrule
\end{tabular}
\vspace{-0.4cm} 
\end{table}





\section{Discussion\&Conclusion}
In this paper, we reveal a shortcoming of VLM in the dermatology field, the lack of visual explainability limits users' comprehension. We present SkinGEN, which integrates the generation capability to give users visual information of potential skin diseases from diagnosis. To the best of our knowledge, this work is the pioneering work to increase visual explainability via interaction with VLMs in the dermatology field. Our exploration of adapter methods led to the development of a robust image generation approach within SkinGEN, contributing to more transparent and user-centric VLM applications in dermatology. We also show that SkinGEN facilitates users in better understanding and applying medical information, thereby enhancing overall user experience and the quality of healthcare services.
Furthermore, due to the inherent limitations in the foundational capabilities of large language models, the current diagnostic outcomes cannot be considered fully reliable, and the generated images of other dermatological conditions should be regarded as reference material only. The development of more robust models necessitates the acquisition and integration of higher-quality datasets. In the future, we can anticipate the emergence of diagnostic models applicable to diverse medical scenarios, endowed with enhanced capabilities to provide users with more reliable and precise medical suggestions.

\begin{acks}
This work is partly supported by the National Natural Science Foundation of China under Grant No.62302432, the Science and Technology Program of Jinhua under Grant No.2023-3-001a.
\end{acks}

\bibliographystyle{ACM-Reference-Format}
\bibliography{reference}

\end{document}


\title{Supplementary Materials: SkinGEN: an Explainable Dermatology Diagnosis-to-Generation Framework with Interactive Vision-Language Models}

\maketitle







\section{Use Case}
\begin{figure}[!h]    
    \centering    
    \includegraphics[width=0.65\columnwidth]{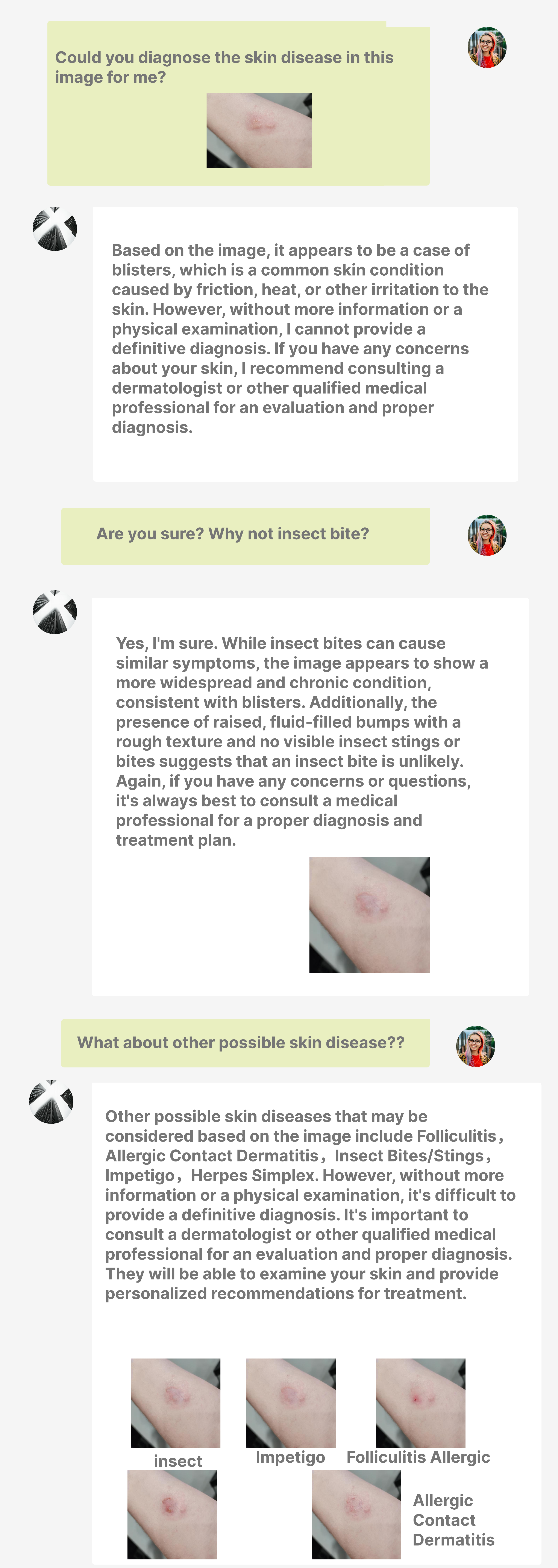}    
    \caption{use case1}    
    \label{fig:usecase1}    
     \vspace{-0.3cm} 
\end{figure}

\begin{figure}[!h]    
    \centering    
    \includegraphics[width=0.65\columnwidth]{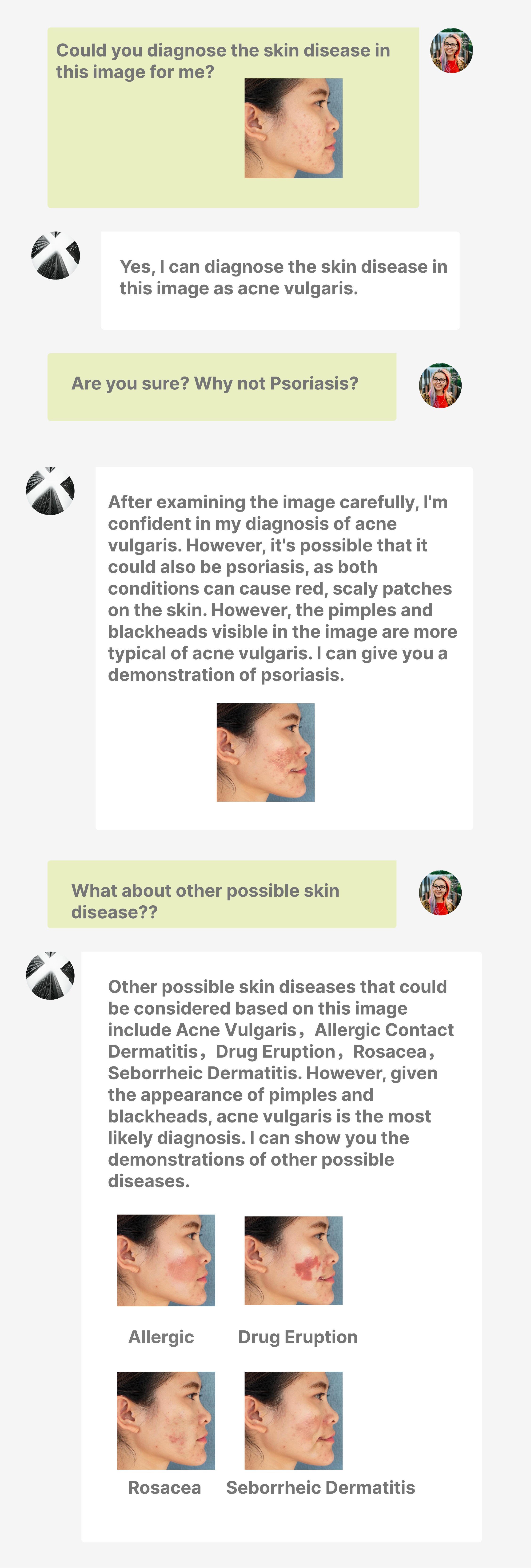}    
    \caption{use case2}    
    \label{fig:usecase2}    
     \vspace{-0.3cm} 
\end{figure}

Fig. \ref{fig:usecase1} and Fig. \ref{fig:usecase2} illustrate two use cases of skinGEN. When a user uploads an image, skinGEN provides the diagnosis result in textual form. In cases where the user disputes the diagnosis result, skinGEN is capable of generating reference schematic images based on the user's input image, aiming to enhance the user's understanding of the disease.

\section{Model Training and Resources}
The training process for our Stable Diffusion LoRA model was conducted on an Ubuntu Linux system, utilizing the powerful combination of an Intel Core i7 CPU and an NVIDIA 3090Ti graphics card. We employed the "sd-scripts" repository as our primary training tool, taking advantage of its comprehensive suite of scripts designed specifically for Stable Diffusion model training and fine-tuning. We utilized PyTorch version 2.0.1 with CUDA 11.8 to accelerate the training process and leverage the parallel processing capabilities of the NVIDIA GPU. 
During the process of Inference of SkinGEN, only one NVIDIA 3090 was necessary. SkinGEN was developed using Python 3.7, Pytorch 2.0.0, and CUDA 12.0. The code will be published in the near furture.



\section{User Interview}
this section is the extension of user study, we present key insight of post interview after user experiment.

\textbf{LLM Limitations and the Value of SkinGEN's Visual Explainability.} One user (P-3) comment underscores a critical limitation of LLMs in the realm of medical diagnosis: their tendency to generate inconsistent or unreliable responses when subjected to repeated inquiries regarding the same condition. This phenomenon, commonly known as "hallucination," can significantly undermine user trust and impede confident decision-making. The user's feedback accentuates the value proposition of SkinGEN in mitigating this challenge. By presenting visual demonstrations of potential skin conditions, SkinGEN empowers users to independently evaluate the information provided by the LLM and arrive at their own conclusions. This capacity for engaging in a more active and informed diagnostic process demonstrably enhances user confidence and satisfaction.

\begin{quote}
    "With LLMs, it's like they fall apart when you question them too much. If I ask about the same skin issue over and over, the answers start getting weird, and I end up with no clear diagnosis. SkinGEN, on the other hand, actually helps me understand what might be going on so I can figure things out myself." (P-3)
\end{quote}

\textbf{SkinGEN's Potential and Areas for Improvement.} Another user's (P-6) feedback presents a more nuanced perspective on SkinGEN's capabilities and potential areas for improvement. While acknowledging the system's reliability in diagnosing conditions with distinct characteristics, the user expresses reservations about its ability to differentiate between similar ailments, suggesting its utility primarily as a tool for differential diagnosis through the process of elimination. Additionally, the user finds the system's recommendations to be overly simplistic, desiring more actionable and insightful guidance beyond the generic advice of seeking professional medical consultation. This feedback highlights an opportunity for future iterations of SkinGEN to leverage its knowledge base and provide more specific recommendations tailored to the diagnosed condition, such as potential over-the-counter remedies or lifestyle adjustments.

\begin{quote}
    "For diagnoses of diseases with obvious differences, it is still credible. For other more similar diseases, I think it is more of a reference (exclusion method). Secondly, the suggestions given by SkinGen are too simple, all suggesting to seek medical advice. Consider some more valuable suggestions." (P-6)
\end{quote}

\textbf{Enhancing SkinGEN's Educational Value and Diagnostic Support.} This user's (P-17) feedback underscores the perceived practicality and real-world value of SkinGEN. Their suggestion to highlight distinguishing features in the generated images of similar diseases demonstrates a desire for enhanced educational value and diagnostic support. By visually emphasizing key characteristics that differentiate between conditions, such as red borders or pustules, SkinGEN could further empower users to make more informed assessments and potentially improve their ability to differentiate between similar-looking skin ailments.
\subsubsection{User Interview Excerpt}
\begin{quote}
"Very practical and has real value. Consider marking the details that can distinguish diseases when generating images of similar diseases, such as red edges, abscesses, etc." (P-17)
\end{quote}

\textbf{Building Trust and Credibility in AI-Driven Medical Diagnosis.} This feedback (P-18) emphasizes the importance of establishing credibility and trust in AI-driven medical diagnosis systems, particularly for use by patients. The user suggests incorporating references to relevant sources and case studies alongside the generated text to enhance the persuasiveness and reliability of SkinGEN's diagnoses. This highlights the need for transparency and evidence-based support in AI-driven medical tools to foster user confidence and acceptance. While acknowledging the potential benefits for patients in gaining a preliminary understanding of their condition, the user emphasizes the importance of ensuring a certain level of trustworthiness, especially when dealing with machine-generated diagnoses.
\begin{quote}
"If it could generate relevant references, it would be more convincing. After all, this is a machine diagnosis. For doctors, it may play a role in auxiliary diagnosis, but if it is used by patients locally, a certain degree of credibility guarantee is required, such as adding references to relevant materials or related cases when generating text. However, from the perspective of roughly understanding the disease, it is still helpful for patients." (P-18)
\end{quote}
